\documentclass[aps,prl,superscriptaddress,showpacs,twocolumn]{revtex4-1}
\usepackage[pdftex]{graphicx}
\usepackage{amsmath}
\usepackage{xspace}
\usepackage{color}
\usepackage[colorlinks=true,linkcolor=blue,urlcolor=blue,citecolor=blue]{hyperref}
\usepackage[normalem]{ulem}
\usepackage{hyperref}

\usepackage{color}
\definecolor{BV}{rgb}{0.1,0.,0.6}
\definecolor{R}{rgb}{0.9,0,0}
\definecolor{G}{rgb}{0.2,0.8,0.2}
\usepackage{comment}

\usepackage{colordvi}

\begin{document}

\title{Reentrant behavior of the breathing-mode-oscillation frequency in a one-dimensional Bose gas}

\author{A. Iu. Gudyma}
\affiliation{Universit\'e Paris-Sud, Laboratoire LPTMS, UMR8626, Orsay, F-91405, France}
\affiliation{CNRS, Orsay, F-91405, France}

\author{G. E. Astrakharchik}
\affiliation{Departament de F\'isica i Enginyeria Nuclear, Campus Nord B4-B5, Universitat Polit\`ecnica de Catalunya, E-08034 Barcelona, Spain}

\author{Mikhail B. Zvonarev}
\affiliation{ Univ Paris-Sud, Laboratoire LPTMS, UMR8626, Orsay, F-91405, France}
\affiliation{CNRS, Orsay, F-91405, France}
\affiliation{ITMO University, 197101, Saint-Petersburg, Russia}

\date{\today}

\begin{abstract}
Exciting temporal oscillations of the density distribution is a high-precision method for probing ultracold trapped atomic gases. Interaction effects in their many-body dynamics are particularly puzzling and counter-intuitive in one spatial dimension (1D) due to enhanced quantum correlations. We consider 1D quantum Bose gas in a parabolic trap at zero temperature and explain, analytically and numerically, how oscillation frequency depends on the number of particles, their repulsion and the trap strength. We identify the frequency with the energy difference between the ground state and a particular excited state. This way we avoided resolving the dynamical evolution of the system, simplifying the problem immensely. We find an excellent quantitative agreement of our results with the data from the Innsbruck experiment [Science \textbf{325}, 1224 (2009)].
\end{abstract}


\maketitle

All existing ultracold-gas experiments are carried out with systems which are spatially inhomogeneous due to the presence of an external confining potential~\cite{pitaevskii_book_BEC,pethick_book_BEC}. Exciting temporal oscillations of the gas density distribution in such a confined geometry is a basic tool for investigating the spectrum of collective excitations and phase diagram~\cite{mewes_breathing_Na_BEC_1996,jin_breathing_96, chevy_breathing_mode_Rb87_2002, kinast_breathing_mode_Li6_2004, kinast_breathing_mode_Li6_2005, altmeyer_breathing_mode_Li6_2007, riedl_breathing_mode_Li6_2008}. One-dimensional (1D) gases have their own specifics: Enhanced quantum correlations affect their collective excitations spectrum drastically, masking out signatures of the Bose and Fermi statistics of the constituent particles~\cite{gogolin_1dbook,giamarchi_book_1d}. The exactly solvable homogeneous Lieb-Liniger gas model~\cite{lieb_boseI_1963} is a paradigmatic demonstration of that statement. There bosons interact through a $\delta$-function potential of strength $g_\mathrm{1D}>0.$ Increasing $g_\mathrm{1D}$ suppresses spatial overlap between any two bosons. This leads to a many-body excitation spectrum identical to that of a free Fermi gas in the limiting case of infinite repulsion, $g_\mathrm{1D}=\infty$, known as the Tonks-Girardeau (TG) gas~\cite{girardeau_impurity_TG_60}. The presence of an external parabolic potential makes the low-lying part of excitation spectrum to be discrete. The first excited state of the gas, a dipole mode, is interaction independent. It is associated with the center-of-mass oscillations at a trap frequency $\omega_z$. The second excited state is doubly degenerate for $g_\mathrm{1D}=0$ and $g_\mathrm{1D}=\infty.$ One mode with the interaction-independent frequency $2\omega_z$ comes from center-of-mass oscillations. Another mode is called the breathing (or compressional) mode. Being excited by a small instantaneous change of the trapping frequency $\omega_z$, this mode has the frequency $\omega$ which depends on $g_\mathrm{1D}>0,$ the number of particles $N$ in the trap, and the gas temperature $T.$
 
Experimental investigations of the breathing mode oscillations in 1D ultracold-gas experiments have been reported by several groups~\cite{moritz_breathing_1D_03,haller_superTonks_2009,fang_breathing_14}. It was found that the frequency ratio $\omega/\omega_z,$ as a function of the interaction strength, goes through two crossovers: from the value $2$ down to $\sqrt{3}$ and then back to $2$  (see, e.g., Fig.~\ref{fig:Hartree}), as the system goes from non-interacting to weakly interacting, and then from weakly interacting to strongly interacting regime~\cite{haller_superTonks_2009}. The latter crossover has been described theoretically for $N$ going to infinity, by the approach based on the local density approximation (LDA)~\cite{menotti_breathing_02}. A description of the former crossover has been done only numerically for the few particles: $N\le 5$ by using the multilayer multiconfiguration time-dependent Hartree method~\cite{schmitz_breathing_few_13} and $N\le 7$ using numerical diagonalization~\cite{tschischik_breathing_13}. Experiments~\cite{moritz_breathing_1D_03,fang_breathing_14} were done in the regime of weak coupling, for which $\omega/\omega_z = \sqrt{3}$ is expected as $N$ goes to infinity at zero temperature. To what extent are the observed deviations from the value $\sqrt 3$ due to finite $N$ and $T$ is an open question. Answering it paves a way towards understanding interaction effects in dynamics and thermalization of 1D quantum gases.

In this Rapid Communication we present the analytic and numerical results for the breathing-mode-oscillation frequency $\omega$ in the repulsive Lieb-Liniger gas in a parabolic trap of frequency $\omega_z$. Using the Hartree approximation we explain how the decrease of $\omega/\omega_z$ from the value $2$ down to $\sqrt 3$ as the interparticle repulsion increases is linked to a transition from the Gaussian Bose--Einstein condensate (BEC) to the Thomas--Fermi (TF) BEC regime. By further increasing the repulsion strength, $\omega/\omega_z$ goes back to the value $2.$ This return is associated with the transition from the TF BEC to the Tonks-Girardeau regime and is described within local density approximation. We perform extensive diffusion Monte Carlo simulations for a gas containing up to $N=25$ particles. As the number of particles increases, predictions from the simulations converge to the ones from the Hartree and LDA in their respective regimes. This makes our results for $\omega$ applicable for arbitrary number of particles and value of the repulsion strength. We find an excellent quantitative agreement with the data from the Innsbruck experiment~\cite{haller_superTonks_2009}. We also estimate relevant temperature scales for the Palaiseau experiment~\cite{fang_breathing_14}.

\begin{figure}
\includegraphics[width=0.98\linewidth,  clip=true,trim= 0 0 0 0]{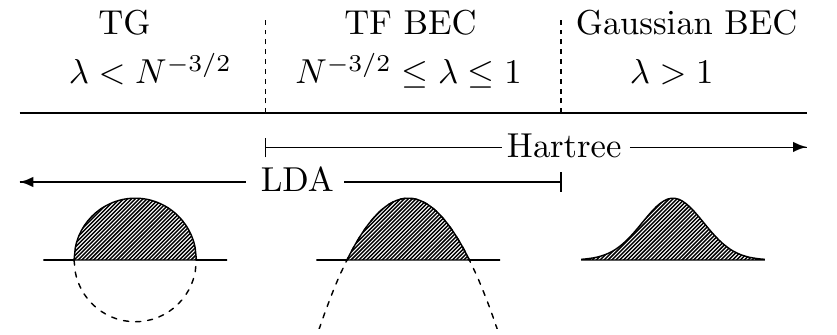}
\caption{The Tonks-Girardeau (TG), the Thomas-Fermi Bose-Einstein condensate (TF BEC), and the Gaussian BEC regimes of the repulsive Lieb-Liniger gas in a parabolic trap, Eq.~\eqref{eq:Ham}, are shown as a function of the Hartree parameter $\lambda = -a_\mathrm{1D}/(N a_z)$ for a given $N.$ Density profiles are semicircle, inverted parabola, and the Gaussian deep in these regimes, respectively. The local density approximation (LDA) parameter $\Lambda$ is related to $\lambda$ as $\Lambda=N^3 \lambda^2.$ The TG and the TF BEC regimes are separated with $\Lambda=1$.}
\label{fig:123}
\end{figure}

\begin{figure}
\centering
\includegraphics[width=0.98\linewidth, clip=true,trim= 0 0 0 0]{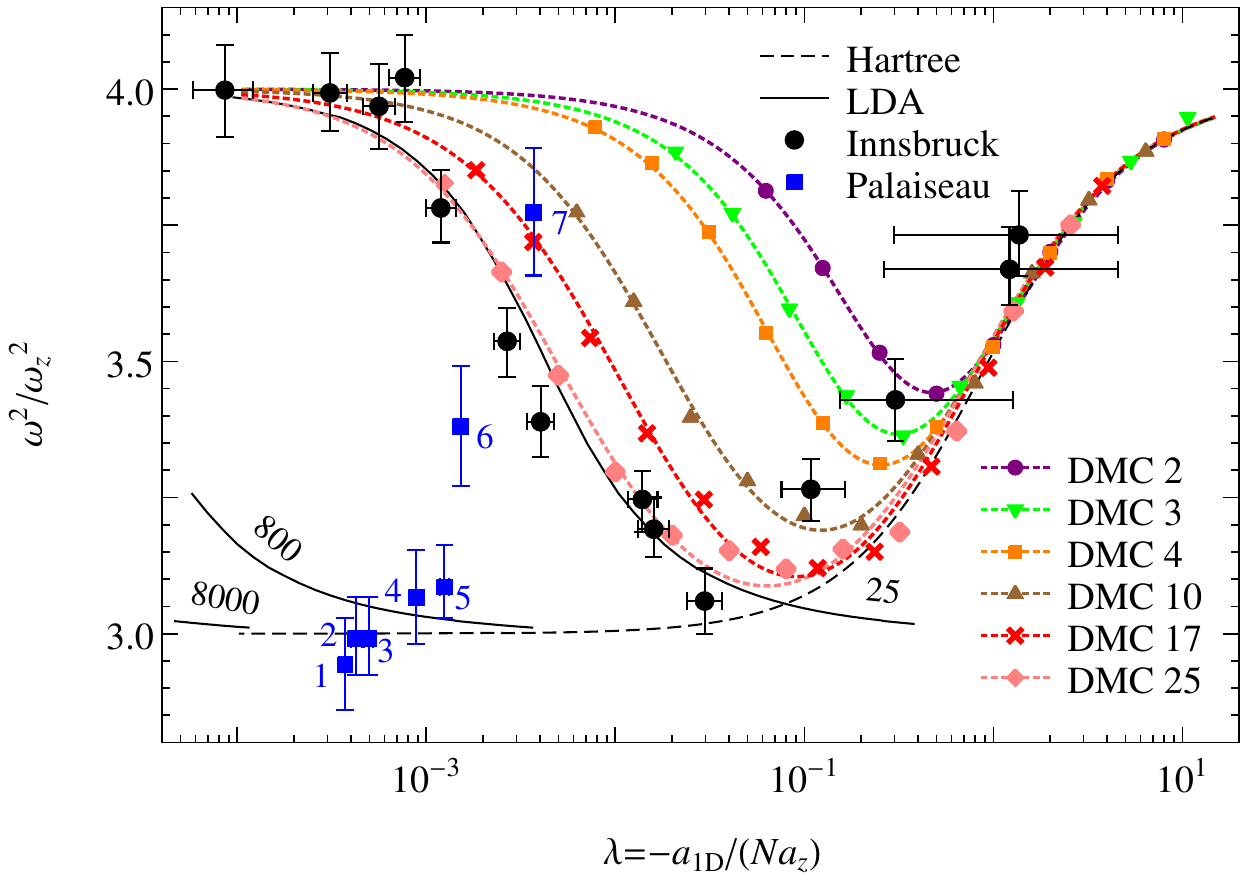}
\caption{(Color online)
Ratio $\omega^2 / \omega_z^2$ as a function of the Hartree parameter $\lambda=-a_\mathrm{1D}/(N a_z)$. Dashed (black) line: the Hartree approximation. Solid (black) lines: LDA for $N=25$, $800$, $8000$ in the equation $\Lambda = N^3 \lambda^2$. Dashed (colored other than in black) lines: interpolations for data points obtained with diffusion Monte Carlo (DMC) simulations for $N=2,3,4,10,17,25$ particles (top to bottom). Large (black) circles: Innsbruck experiment~\cite{haller_superTonks_2009}, for which $N=25$. Large (blue) boxes: Palaiseau experiment~\cite{fang_breathing_14}, for which $N$ is given in Table~\ref{tab:palaiseau}. The Gaussian BEC regime corresponds to $\lambda> 1,$ as defined in Fig.~\ref{fig:123}.
}
\label{fig:Hartree}
\end{figure}

\textit{Model and sum rules}. --- The model we consider is the Lieb-Liniger gas of repulsive bosons in a parabolic trap. The Hamiltonian for $N$ particles is
\begin{equation}
H = -\frac{\hbar^2}{2m} \sum_{i=1}^N \frac{\partial^2}{\partial z_i^2} + g_\mathrm{1D} \sum_{i<j} \delta(z_i-z_j) + \sum_{i=1}^N V(z_i). \label{eq:Ham}
\end{equation}
Here $m$ is the particle mass, and $V(z)=m\omega_z^2 z^2/2$ is the particle potential energy in a parabolic trap. The length scales in the model are set by the $s$-wave scattering length $a_\mathrm{1D}$, related to the coupling constant $g_\mathrm{1D} = -2\hbar^2/(ma_\mathrm{1D})$ and by the harmonic oscillator length $a_z=\sqrt{\hbar/(m\omega_z)}$. Three zero-temperature quantum regimes shown schematically in Fig.~\ref{fig:123} are identified for model~\eqref{eq:Ham} based on its thermodynamic and local correlation properties~\cite{petrov_QG_lectures_04}.

We employ a sum rule approximation, which makes it possible to get $\omega$ for arbitrary $a_\mathrm{1D}$, $a_z$, and $N$ from ground-state properties of Hamiltonian~\eqref{eq:Ham} solely~\cite{abraham_breathing_mode_review_14}. More specifically, $\omega$ is obtained by calculating the response of the gas to a change of the trap frequency:
\begin{equation}
\omega^2= -2 \frac{\langle Q \rangle}{\partial\langle Q\rangle/\partial\omega_z^2},  \label{eq:omegasr:finite size}
\end{equation}
where $Q = Q_c \equiv \sum_{i=1}^N  (z_i-Z_\mathrm{cm})^2$, and $Z_\mathrm{cm} = \sum_{i=1}^N z_i / N$ is the center-of-mass coordinate. The average $\langle \cdots \rangle $ is taken with respect to the ground-state wave function $\psi_\mathrm{gs}(z_1,\ldots,z_N)$ of Hamiltonian~\eqref{eq:Ham}. Neglecting $Z_\mathrm{cm}$ in $Q_c$ amounts to replacing $Q_c$ with $ Q_0 \equiv \sum_{i=1}^N z_i^2$, the latter operator being used in Ref.~\cite{menotti_breathing_02}.

By changing $\omega_z$ one excites many modes rather than a single breathing mode. These modes cause $\omega$ given by Eq.~\eqref{eq:omegasr:finite size} to be different from the breathing mode frequency. Their contribution could be diminished by a proper choice of $Q.$ How good is our choice, $Q_c,$ for that purpose is seen by comparing the exact spectrum of model~\eqref{eq:Ham} for $N=2$
with $\omega$ given by Eq.~\eqref{eq:omegasr:finite size} for arbitrary value of $-a_\mathrm{1D}/a_z$. We found that $\omega$ given by Eq.~\eqref{eq:omegasr:finite size} with $Q=Q_0$ misses up to $50\%$ of the deviation from $2\omega_z$ value, while with $Q=Q_c$ it misses $4\%$ at most.

\textit{Gaussian BEC to TF BEC crossover}. --- We approximate $\psi_\mathrm{gs}$ (normalized to $\langle\psi_\mathrm{gs}|\psi_\mathrm{gs}\rangle =1$) with the Hartree variational wave function $\psi_\mathrm{gs}^H (z_1,\ldots,z_N) = \prod_{i=1}^N \varphi(z_i)$ for $N\gg 1.$ This function is found by minimizing the functional $E[\psi_\mathrm{gs}^H]\equiv \langle\psi_\mathrm{gs}^H|H|\psi_\mathrm{gs}^H\rangle$ with respect to $\varphi(z).$ The procedure amounts to solving the Hartree eigenvalue equation (same as the Gross-Pitaevskii equation)
\begin{equation}
\left[-\frac12\frac{\partial^2}{\partial x^2} + \frac{x^2}2 + \frac2{\lambda} |\tilde\varphi(x)|^2 \right] \tilde\varphi(x) = \epsilon \tilde\varphi(x) \label{eq:Heq}
\end{equation}
for the minimal possible $\epsilon$. Here $\tilde\varphi(x)=\sqrt{a_z}\varphi(z),$ and $x$ and $\epsilon$ are dimensionless length and energy given in units of $a_z$ and $\hbar\omega_z,$ respectively. The Hartree parameter $\lambda$ reads
\begin{equation}
\lambda = -\frac{a_\mathrm{1D}}{Na_z}.
\label{Eq:lambda}
\end{equation}
The ground-state density distribution found with respect to the Hartree state $|\psi_\mathrm{gs}^H\rangle$ is $n_H(z)=N|\varphi(z)|^2$ and the average of the operators $Q_c$ and $Q_0$ is $\langle Q_c \rangle = \langle Q_0 \rangle = \int dz\, z^2 n_H(z)$. Substituting this expression into Eq.~\eqref{eq:omegasr:finite size} and taking into account that $2\partial\lambda/\partial\omega_z = \lambda/\omega_z $ we find that $\omega/\omega_z$ depends on $a_\mathrm{1D}$, $a_z$, and $N$ through a single parameter $\lambda$ within the Hartree approximation.

We explore the dependence of $\omega/\omega_z$ on $\lambda.$ Deep in the Gaussian BEC regime, $\lambda\gg 1$, we use a series expansion in the harmonic oscillator wave functions for $\varphi(z)$ and solve Eq.~\eqref{eq:Heq} perturbatively. We get
\begin{equation}
\omega^2/\omega_z^2 \simeq 4 \left(1 - c \lambda^{-1}\right), \quad \lambda\to\infty, \label{eq:RHexpand}
\end{equation}
where $c=1/\sqrt{8\pi}$. Perturbation theory for the many-body wave functions of Hamiltonian~\eqref{eq:Ham} extends the validity range of Eq.~\eqref{eq:RHexpand} to arbitrary $N\ge 2.$ Note that the Hartree approximation is only valid in the large $N$ limit. Indeed, Eq.~\eqref{eq:Heq} in which $\lambda$ is replaced with $\lambda_N = -a_\mathrm{1D}/[(N-1)a_z]$ minimizes the energy functional for any $N\ge 2$. This implies $ \left< Q_0 \right>  = N \left< Q_c \right> /(N-1) = \int dz\, z^2 n_H(z).$ Being substituted into Eq.~\eqref{eq:omegasr:finite size} both $ \left< Q_0 \right> $ and $ \left< Q_c \right> $ lead to Eq.~\eqref{eq:RHexpand} with $\lambda$ replaced by $\lambda_N,$ that is, to the result which is correct in the large $N$ limit only. Note also that the ground-state wave function of Hamiltonian~\eqref{eq:Ham} obtained with perturbation theory and used for the sum rule~\eqref{eq:omegasr:finite size} with $Q=Q_c$ gives Eq.~\eqref{eq:RHexpand} correctly. This supports our approach to DMC simulations (detailed later in the Rapid Communication), in which we rely on Eq.~\eqref{eq:omegasr:finite size} and $Q = Q_c$.

In the case $\lambda\ll 1$ Eq.~\eqref{eq:Heq} results in an inverted parabola density profile, characteristic of the TF BEC regime
\begin{equation}
n_H(z) = N\frac{(9\lambda)^{\frac13}}{4a_z} \left(1-\frac{z^2}{Z^2}\right) \theta\left(1-\frac{z^2}{Z^2}\right), \quad \lambda\to0. \label{eq:nHartree_TFBEC}
\end{equation}
Here $\theta$ is the Heaviside step function, and $Z/a_z = \left(3/\lambda\right)^{1/3}$. Substituting Eq.~\eqref{eq:nHartree_TFBEC} into Eq.~\eqref{eq:omegasr:finite size} we get $\omega/\omega_z = \sqrt{3}$.

In the case of arbitrary $\lambda$ we solve Eq.~\eqref{eq:Heq} numerically. The plot of $\omega^2/\omega_z^2$ as a function of $\lambda$ is shown in Fig.~\ref{fig:Hartree}. We observe a smooth crossover between the $\lambda\ll 1$ (TF BEC) and $\lambda\gg 1$ (Gaussian BEC) regimes.  We find that $\omega^2/\omega_z^2 \approx 3.5$ at $\lambda=1,$ defined as a reference point separating these regimes (see Fig.~\ref{fig:123}).

\textit{TF BEC to TG crossover}. --- This crossover is associated to an interplay of the parameters $\lambda$ and $N^{-3/2}$; see Fig.~\ref{fig:123}. It may not be captured within the Hartree approximation, which does not contain $N^{-3/2}$ as a parameter independent of $\lambda$. Instead, we may use LDA~\cite{menotti_breathing_02}. It is only valid in the large $N$ limit, and is based on the assumption that the local chemical potential at a point $z$ is equal to the chemical potential in a homogeneous system that has the same density $n(z).$ Therefore $\mu_\mathrm{loc}(n(z))=V(Z)-V(z)$ for $|z|\le Z$ and vanishes for $|z|>Z$ in model~\eqref{eq:Ham}. Here $Z$ is the Thomas-Fermi radius of the gas cloud, whose value is set by the normalization condition $\int_{-Z}^Z dz\, n(\mu_\mathrm{loc}(z))=N.$  The dependence of $\mu_\mathrm{loc}$ on $n$ in the homogeneous Lieb-Liniger model (Eq.~\eqref{eq:Ham} with $V=0$) was found in Ref.~\cite{lieb_boseI_1963}. Using Eq.~\eqref{eq:omegasr:finite size} with $\langle Q_c \rangle = \langle Q_0 \rangle = \int dz\, z^2 n(z)$ we get $\omega/\omega_z$ readily. The result depends on $a_\mathrm{1D}, a_z,$ and $N$ through a single parameter $\Lambda = N a_\mathrm{1D}^2 / a_z^2 = N^3 \lambda^2$ within LDA~\cite{petrov_DegQGRegimes1D_00, dunjko_DegQGRegimes1D_01, menotti_breathing_02}. 

In the limiting case of impenetrable bosons, $\Lambda =0$, the local chemical potential is equal to the Fermi energy, $\mu_\mathrm{loc} = (\pi\hbar n)^2/(2m).$ This leads to the semicircular LDA density profile $n(z)= \sqrt{2N a_z^2-z^2} \theta(2N a_z^2 -z^2)/(\pi a_z^2)$, characteristic of the TG regime, see Fig.~\ref{fig:123}. Excitation spectrum of model~\eqref{eq:Ham} deep in the TG regime, $\Lambda \ll 1,$ can be found perturbatively in $a_\mathrm{1D}$. For that we use a mapping from the gas of strongly repulsive bosons to that of weakly attractive fermions~\cite{cheon_fermionboson_dual_1D_99}. A perturbative solution for the ground state energy is given in Ref.~\cite{paraan_pertirbation_TG_trapped_1D_10}. Analyzing the excited states results in the expansion~\cite{zhang_scale_breakdown_TG_14}
\begin{equation}
\omega^2/\omega_z^2 \simeq 4 \left(1 - C_N \sqrt{\Lambda} \right), \quad \Lambda\to 0, \label{eq:RTGexpand}
\end{equation}
where $C_N$ is calculated for all $N\ge 2:$
\begin{multline}
C_N = \frac{3 \sqrt{2 N}}{\pi \sqrt{\pi}} \frac{\Gamma(N-\frac{5}{2}) \Gamma(N+\frac{1}{2})}{\Gamma(N) \Gamma(N+2)}\\
\times {}_3 F_2 \left(\frac{3}{2},1-N,-N;\frac{7}{2}-N,\frac{1}{2}-N;1 \right).
\label{eq::TG_expansion}
\end{multline}
The $\Lambda\to 0$ expansion of the LDA solution reproduces Eq.~\eqref{eq:RTGexpand} and the coefficient $C_\infty.$ Note that the coefficient $c$ entering Eq.~\eqref{eq:RHexpand} does not depend on $N,$ while $C_N$ grows monotonously from $C_2=1/\sqrt{4 \pi} \approx 0.282$ to $C_\infty = 32\sqrt{2}/(15\pi^2) \approx 0.306.$

In the case $\Lambda\gg 1$, local chemical potential is of the Gross--Pitaevskii form, $\mu_\mathrm{loc}=g_\mathrm{1D}n,$ and the shape of the density profile is given by Eq.~\eqref{eq:nHartree_TFBEC}, characteristic of the TF BEC regime. This implies $\omega / \omega_z = \sqrt{3}.$

\begin{figure}.
\centering
\includegraphics[width=0.98\linewidth, clip=true,trim= 0 0 0 0]{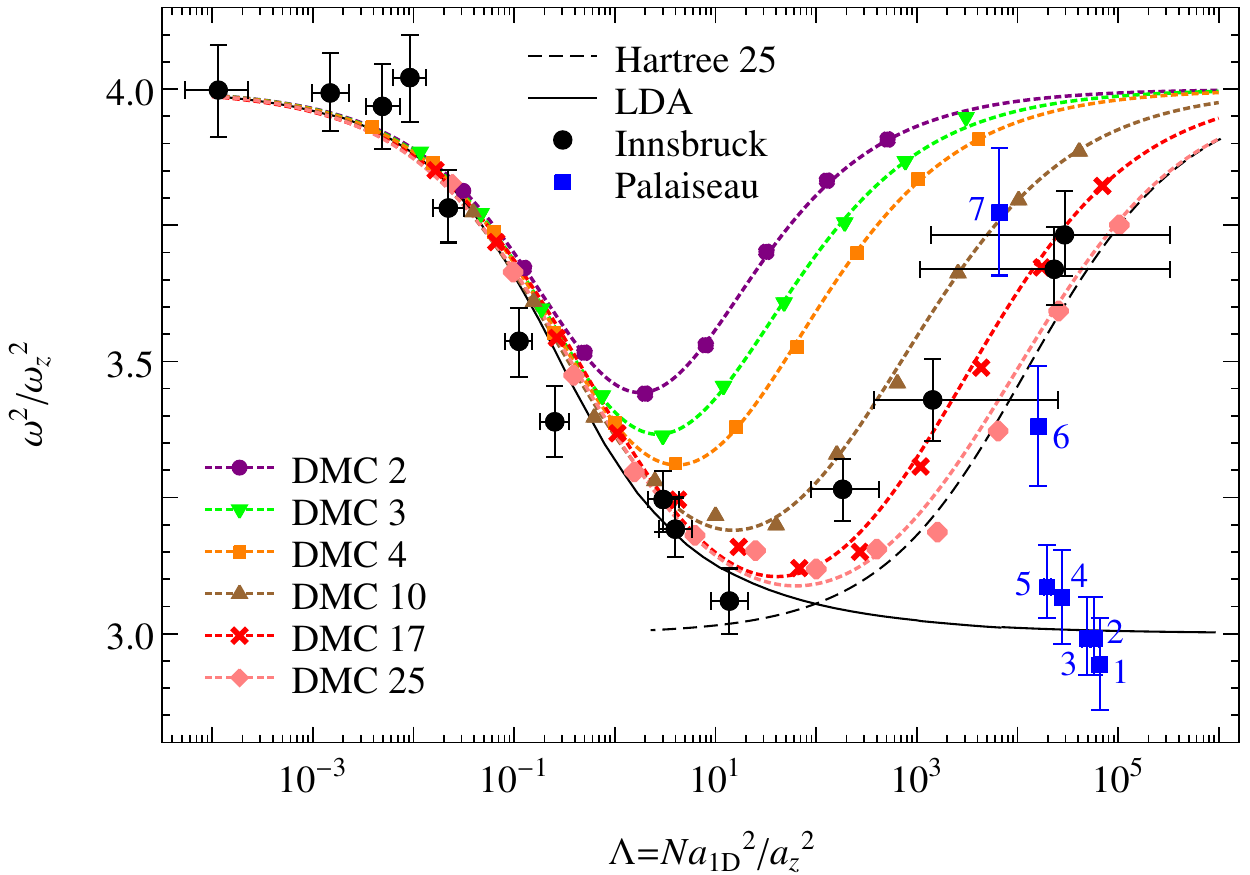}
\caption{(Color online)
Ratio $\omega^2/\omega_z^2$, as a function of LDA parameter $\Lambda= N a_\mathrm{1D}^2/a_z^2$. Dashed (black) line: the Hartree approximation for $N=25$ in the equation $\Lambda = N^3 \lambda^2$. Solid (black) line: LDA. DMC and experimental data points are the same as in Fig.~\ref{fig:Hartree}. The TG regime corresponds to $\Lambda < 1,$ as defined in Fig.~\ref{fig:123}.
}
\label{fig:LDA}
\end{figure}

In the case of arbitrary $\Lambda$ we solve the Lieb's integral equations connecting $\mu_\mathrm{loc}$ and $n$ numerically. We see from Fig.~\ref{fig:LDA} that $\omega / \omega_z$ connects smoothly $\Lambda\ll 1$ (TG) and $\Lambda\gg 1$ (TF BEC) regimes~\cite{menotti_breathing_02}.
We find that $\omega^2/\omega_z^2 \approx 3.3$ at $\Lambda=1,$ defined as a reference point separating these regimes (see Fig.~\ref{fig:123}).

\textit{DMC simulations}. --- How does $\omega/\omega_z$ depend on model parameters for small $N$, and how good are the Hartree approximation/LDA in that case? To answer these questions quantitatively we perform large-scale numerical simulations based on the diffusion Monte Carlo (DMC) algorithm~\cite{boronat_Aziz_QMC_94}. This algorithm amounts to solving many-body Schr\"odinger equation in imaginary time and makes it possible to calculate ground-state energy to arbitrarily high precision. The convergence rate of the simulations can be enhanced greatly by doing an importance sampling with a guiding wave function $\psi_T$. We use $\psi_{T}(z_1,...,z_N) = \prod_{i=1}^N \exp(-c_\mathrm{var} z_i^2)\prod_{j<k}^N (|z_j-z_k|-a_\mathrm{1D}),$ with the parameter $c_\mathrm{var}$ minimizing the variational energy. This function is known to work very well in a number of 1D systems~\cite{astrakharchik_Lieb_QMC_03, astrakharchik_Lieb_QMC_06, astrakharchik_fermi_impurity_13, garcia-march_fewatom_crossover_13}.

We use the sum rule approximation~\eqref{eq:omegasr:finite size}, which only requires the knowledge of the ground state properties of the model. For the number of particles ranging from $N=2$ to $25$ we pushed DMC to its limits to perform high-accuracy simulations. Specifically, up to $10^4$ CPU hours were used to get each data point for $\omega/\omega_z$. The results obtained are shown as a function of $\lambda$ in Fig.~\ref{fig:Hartree} and of $\Lambda$ in Fig.~\ref{fig:LDA}. Dashed lines interpolating the data points are obtained by using a Pad\'e approximation and Eqs.~\eqref{eq:RHexpand} and~\eqref{eq:RTGexpand} for the asymptotic values of $\omega/\omega_z.$ We see in Fig.~\ref{fig:Hartree} that the Hartree and DMC curves are indistinguishable from each other for $\lambda > 1$ at any $N$. The minimal value of $\lambda$ at which these two curves are close to each other decreases with increasing $N.$ It reaches the value $\approx 0.1,$ and the minimal value of $\omega^2/\omega^2_z$ reaches $ \approx 3.2,$ at $N=25.$ We may thus locate the TF BEC regime of the model from Fig.~\ref{fig:Hartree} by setting where $\omega^2/\omega^2_z \approx 3$. Figure~\ref{fig:LDA} shows the same data points as in Fig.~\ref{fig:Hartree}, as a function of the LDA parameter~$\Lambda$. Evidently, LDA and DMC curves coincide for $\Lambda< 0.1$ at any $N$.

\textit{Comparison with experiments}. --- The Innsbruck group loaded three-dimensional (3D) BEC of $\vphantom{}^{133}\textrm{Cs}$ atoms into an array of 1D tubes formed by retro-reflected laser beams. The frequency of the external parabolic potential along the tube direction is $\omega_z=2\pi \times 15.4$ $\mathrm{Hz},$ and the maximal number of atoms per tube is about $25.$ The Innsbruck group data shown in Figs.~\ref{fig:Hartree} and \ref{fig:LDA} of the present Rapid Communication are taken from Figs.~2 and 3(a) of Ref.~\cite{haller_superTonks_2009} for $g_\mathrm{1D}>0$. We see that DMC simulations for $N=17$ and $25$ are compatible with the experimental data points. This match suggests that the temperature effects play little role in the experiment. The temperature $T$ of the 1D gas can be estimated by assuming that it is inherited from the 3D BEC, whose temperature is between $1$ and $10$ $\mathrm{nK}$~\cite{Note1}. The degeneracy temperature of an ideal Bose gas, defined as $T_Q=N\hbar\omega_z/k_B$ ($k_B$ is the Boltzmann constant) is about $18$ $\mathrm{nK}.$ We see that $T$ is at least twice as low as $T_Q.$

The ETH experiment examined what happens with the breathing oscillations if the temperature of the 3D BEC prepared to be loaded into an array of 1D tubes gets higher~\cite{moritz_breathing_1D_03}. The parameters $a_\mathrm{1D},$ $a_z,$ and $N$ correspond to the TF BEC regime of the 1D gas. It was found that the breathing mode persists and $\omega^2/\omega_z^2$ grows from the value $3$ to $4$ (with the uncertainty about $0.1$). These findings could be interpreted as the increase of $\omega$ due to the increase of the temperature of the 1D gas, assuming that it is in thermal equilibrium.

The Palaiseau group prepared a single tube with $\vphantom{}^{87}\mathrm{Rb}$ atoms using atom-chip setup~\cite{fang_breathing_14}. The number of atoms in the tube is given in Table~\ref{tab:palaiseau}, and $\omega_z=2\pi \times 9.0 $ $\mathrm{Hz}.$ Data points from the Palaiseau group shown in Figs.~\ref{fig:Hartree} and \ref{fig:LDA} of the present Rapid Communication are taken from Fig.~3(a) of Ref.~\cite{fang_breathing_14}. The parameters $a_\mathrm{1D},$ $a_z,$ and $N$ correspond to the TF BEC regime for all data points. We see that the frequencies for the first five of them match our theoretical predictions within the error bars. The frequencies for the last two of them are higher than the theory predicts.
\begin{table}
\begin{ruledtabular}
\begin{tabular}{lccccccc}
Point & $1$ & $2$ & $3$ & $4$ & $5$ & $6$ & $7$ \\ \colrule
$N\times 10^{-3}$ & $7.8$ & $6.8$ & $ 5.8$ & $ 3.2$ & $ 2.3$ & $ 1.9$ & $ 1.4$ \\
$n_0$ $[\mathrm{\mu m}^{-1}]$ & 66 & 58 & 52 & 33 & 23 & 18 & 13 \\
$\omega^2/\omega_z^2$ & 2.94 & 2.99 & 2.99 & 3.07 & 3.09 & 3.38 & 3.77 \\
$T$ $[\mathrm{\mu K}]$ & 0.40 & 0.40 & 0.34 & 0.23 & 0.21 & 0.21 & 0.19 \\
$T_\mathrm{co}$ $[\mathrm{\mu K}]$ & 1.08 & 0.95 & 0.82 & 0.50 & 0.38 & 0.33 & 0.22
\end{tabular}
\end{ruledtabular}
\caption{Comparison with  the data for the Palaiseau experiment.  The number of particles $N$ in the tube and the density $n_0$ at the tube center are from the raw data used in Ref.~\cite{fang_breathing_14}. The ratio $\omega^2/\omega_z^2$ is taken from Fig.~3(a) of Ref.~\cite{fang_breathing_14} and is shown as the large (blue) boxes in Figs.~\ref{fig:Hartree} and \ref{fig:LDA} of the present Rapid Communication. The temperature $T$ is obtained by requiring that the height of the thermal gas density profile at the trap center is equal to $n_0$. The parameter $T_\mathrm{co}$ determines when the finite-temperature effects are important according to Ref.~\cite{bouchoule_interaction_crossover_07}.
\label{tab:palaiseau}}
\end{table}
We get the gas temperature by comparing the height of the density profile in the tube center, $n_0,$ calculated theoretically~\cite{Note2} with the one measured in experiment~\cite{fang_breathing_14}. The values of $n_0$ and $T$ are given in Table.~\ref{tab:palaiseau}. According to Ref.~\cite{bouchoule_interaction_crossover_07}, finite-temperature effects are relevant above $T_\mathrm{co}= 3N\hbar\omega_z/[k_B\ln(\Lambda/4)]$ for the range of parameters chosen in the experiment. We see from Table~\ref{tab:palaiseau} that $T/T_\text{co} $ increases monotonously from the value $\approx 0.4$ for the first data point to $\approx 0.9$ for the last one. Note that $T_Q$ is nearly three times larger than $T_\text{co}$ (and, therefore, than $T$) for all data points. Thus, $T_Q$ may not define the crossover temperature in the experiment~\cite{fang_breathing_14}.

\textit{Summary}. --- We investigated the breathing mode frequency $\omega$ in model~\eqref{eq:Ham} at zero temperature by identifying the energy difference between a particular excited state and the ground state. This way we avoided dealing with the dynamical evolution of the initial state of the system. Our theory predicts the reentrant behavior of $\omega$ and fully explains the recent experiment~\cite{haller_superTonks_2009} for the repulsive interparticle interaction. The extension of the present theory to the finite-temperature case requires a separate study. The existing phenomenological approaches~\cite{fang_breathing_14,hu_collective_modes_1DBose_finiteT_2014} are yet to  be tested against the predictions from the exact dynamical evolution of the system.

\acknowledgments
We thank the Palaiseau group~\cite{fang_breathing_14} for providing access to their experimental data and the Innsbruck group~\cite{haller_superTonks_2009} for numerous enlightening discussions. The Barcelona Supercomputing Center (The Spanish National Supercomputing Center -- Centro Nacional de Supercomputaci\'on) is acknowledged for the provided computational facilities. The work of A.Iu.G. was supported by grant from Region Ile-de-France DIM NANO-K. G.E.A. acknowledges partial financial support from the DGI (Spain) Grant No.~FIS2011-25275 and Generalitat de Catalunya Grant No.~2009SGR-1003.


%

\end{document}